\documentclass[twocolumn]{autart}    

\usepackage{graphicx}          
\usepackage{booktabs}       
\usepackage{wrapfig}
\usepackage{appendix}
\usepackage{float}
\usepackage[fleqn]{amsmath}
\usepackage{graphicx}
\usepackage{amsmath}
\usepackage{amsbsy}
\usepackage{amssymb}
\usepackage{latexsym}
\usepackage{color}
\usepackage{wrapfig}
\usepackage{algorithmic}

\let\theoremstyle\relax
\usepackage{amsthm}

\newtheorem{proposition}{Proposition}

\newtheorem{definition}{Definition}

\newtheorem{theorem}{Theorem}

\newtheorem{assumption}{Assumption}

\newtheorem{remark}{Remark}

\newcommand{\signal}[1]{{\boldsymbol{#1}}}
\newcommand{\norm}[1]{\left\|#1\right\|}
\newcommand{\Real}{{\mathbb R}}
\newcommand{\integer}{{\mathbb Z}}
\newcommand{\refeq}[1]{(\ref{#1})}
\newcommand{\reftab}[1]{Table \ref{#1}}
\newcommand{\reffig}[1]{Figure \ref{#1}}
\newcommand{\argmin}{\operatornamewithlimits{argmin}}

\newcommand{\X}{{\mathcal{X}}}

\newcommand{\U}{\mathcal{U}}
\newcommand{\T}{{\sf T}}
\newcommand{\mathO}{\mathcal{O}}
\newcommand{\x}{\times}
\newcommand{\inpro}[1]{\left<#1\right>}

\definecolor{darkgreen}{rgb}{0,.6,0}
\definecolor{medorange}{rgb}{0.7,0.3,0}
\definecolor{cyancyan}{rgb}{0.68, 0.92, 0.92}
\def\nn{\nonumber}

\begin{document}

\begin{frontmatter}

\title{Constraint Learning for Control Tasks with \\Limited Duration Barrier Functions\thanksref{footnoteinfo}} 

\thanks[footnoteinfo]{Corresponding author M. Ohnishi.  (Paul G. Allen School of CS \& E) Tel. +1 206-543-1695. Fax +1 206-543-2969.}

\author[RIKEN,UW]{Motoya Ohnishi}\ead{mohnishi@cs.washington.edu},    
\author[GAmech]{Gennaro Notomista}\ead{g.notomista@gatech.edu},               
\author[RIKEN,UT]{Masashi Sugiyama}\ead{sugi@k.u-tokyo.ac.jp}  
\author[GAece]{Magnus Egerstedt}\ead{magnus@gatech.edu}  

\address[RIKEN]{RIKEN Center for Advanced Intelligence Project\\Tokyo 103-0027, Japan}  
\address[UW]{Paul G. Allen School of Computer Science and Engineering, University of Washington\\
	Seattle, WA 98195 USA}             
\address[GAmech]{School of Mechanical Engineering, Georgia Institute of Technology\\
 Atlanta, GA 30313 USA}        
\address[UT]{Department of Complexity Science and Engineering, University of Tokyo\\
	Chiba 277-8561, Japan}   
\address[GAece]{School of Electrical and Computer Engineering, Georgia Institute of Technology\\
	 Atlanta, GA 30332 USA}        


\begin{abstract}                          
	When deploying autonomous agents in unstructured environments over sustained periods of time, adaptability and robustness oftentimes outweigh optimality as a primary consideration.  In other words, safety and survivability constraints play a key role and in this paper, we present a novel, constraint-learning framework for control tasks built on the idea of constraints-driven control.  However, since control policies that keep a dynamical agent within state constraints over infinite horizons are not always available, this work instead considers constraints that can be satisfied over some finite time horizon $T > 0$, which we refer to as limited-duration safety.  Consequently, value function learning can be used as a tool to help us find limited-duration safe policies.  We show that, in some applications, the existence of limited-duration safe policies is actually sufficient for long-duration autonomy.  This idea is illustrated on a swarm of simulated robots that are tasked with covering a given area, but that sporadically need to abandon this task to charge batteries.  We show how the battery-charging behavior naturally emerges as a result of the constraints.  Additionally, using a cart-pole simulation environment, we show how a control policy can be efficiently transferred from the source task, balancing the pole, to the target task, moving the cart to one direction without letting the pole fall down.
\end{abstract}

\end{frontmatter}

\section{Introduction}
\label{sec:introduction}
Acquiring an optimal policy that attains the maximum return over some time horizon
is of primary interest in the literature of both reinforcement learning \cite{reinforcement} and optimal control \cite{optimalcontrol}.
A large number of algorithms have been designed to successfully control systems with complex dynamics
to accomplish specific tasks optimally in some sense.
As we can observe in the daily life, on the other hand, it is often difficult to attribute optimality to human behaviors (cf. \cite{skinner1953science}).
Instead, humans are capable of generalizing the behaviors acquired through completing a certain task
to deal with unseen situations.  This fact casts a question of how one should design a learning algorithm
that generalizes across tasks.

In this paper, we hypothesize that this can be achieved by letting agents acquire a set of {\em good enough} policies when completing one task, and reuse this set for another task.
Specifically, we consider safety, which refers to avoiding certain states, as useful information
shared among different tasks, and we regard limited-duration safe policies as good enough policies (Definition \ref{defge}). 
Our work is built on the idea of constraints-driven control \cite{egerstedt2018robot,lozano2014constraint}, a methodology for controlling agents through enforcement of state constraints.

\begin{figure}[t]
	\begin{center}
		\includegraphics[clip,width=0.35\textwidth]{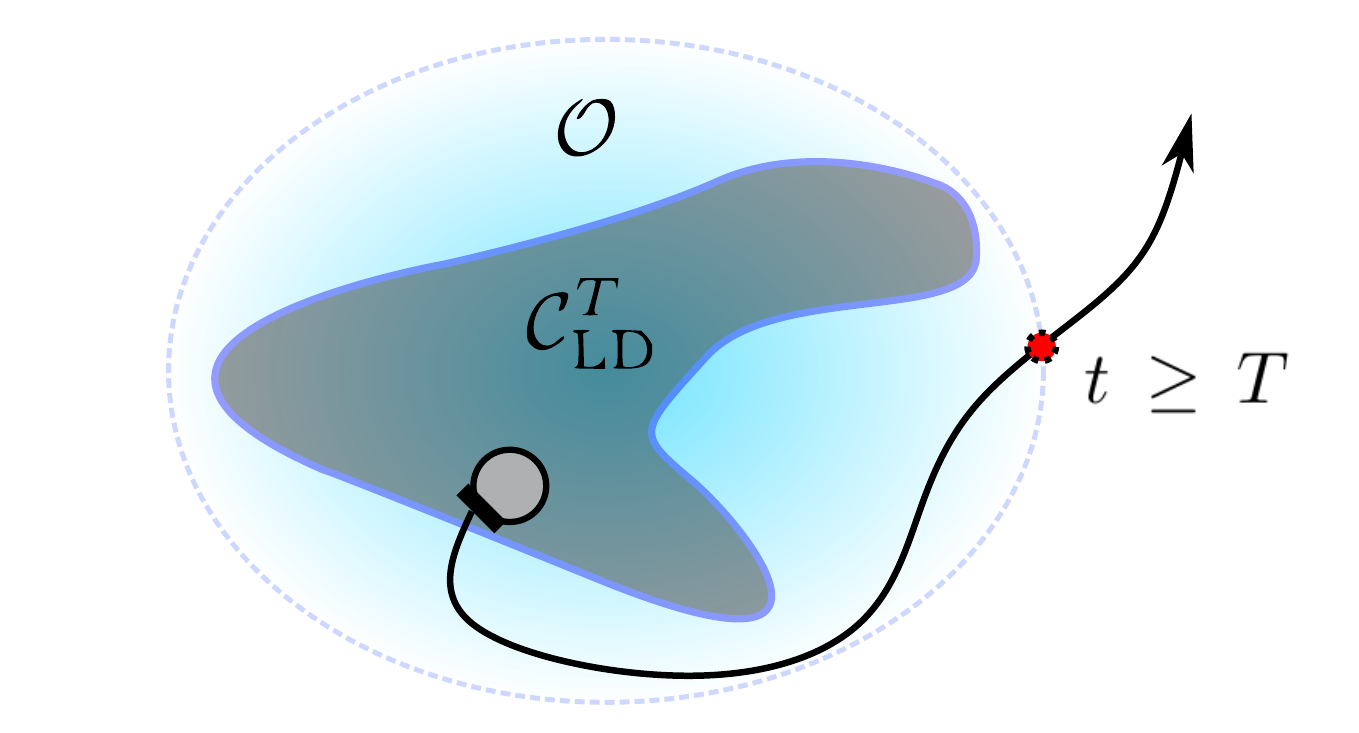}
		\caption{An illustration of limited-duration safety.  An agent stays in $\mathO$ for all $0\leq t< T$ whenever starting from inside the set $\mathcal{C}^T_{\rm LD}\subset\mathO$.}
		\label{fig:SLD}
	\end{center}
\end{figure}

However, state constraints cannot be always satisfied over an infinite-time horizon.
We tackle this feasibility issue by relaxing safety to {\em limited-duration safety},
by which we mean satisfaction of safety over some finite time horizon $T>0$ (see \reffig{fig:SLD}).
To guarantee limited-duration safety, we propose a limited duration control barrier function (LDCBF).  The idea is based on local, model-based control that constrains the instantaneous control input every time to restrict the growths of values of LDCBFs by solving a quadratic programming (QP).

To find an LDCBF, we make use of value function learning, and show that the value function associated with any given policy becomes an LDCBF (Section \ref{subsec:finding}).
Contrary to the optimal control approaches that only single out an optimal policy, 
our framework can be contextualized within the so-called lifelong learning \cite{thrun1995lifelong} and transfer learning \cite{pan2010survey}; see Section \ref{subsec:app2}).

The rest of this paper is organized as follows: Section \ref{sec:related} discusses the
related work, Section \ref{sec:pre} presents
notations, assumptions made in this paper, and some background knowledge.
Subsequently, we present our main contributions and their applications, including simulated experiments, in Section \ref{sec:main} and \ref{sec:app}, respectively.

\section{Related Work}
\label{sec:related}
Finding feasible control constraints that translate to state constraints has been of particular interest
both in the controls and machine learning communities.
Early work includes the construction of navigation functions in terms of obstacle avoidance \cite{rimon1992exact}.
Alternatively, existence of a control Lyapunov function (CLF) \cite{sontag1989universal} enables stabilization of the system, and
CLFs may be learned through demonstrations \cite{khansari2014learning}.
As inverse optimality \cite{freeman1996inverse} dictates that a stabilizing policy is equivalent to an optimal policy in terms of
some cost function, these approaches may be viewed as optimization-based techniques.

On the other hand, control barrier functions (CBFs) \cite{xu2015robustness,wieland2007constructive,ames2017control,glotfelter2017nonsmooth,ohnishi2018safety,ohnishi2018continuous} were proposed to guarantee forward invariance \cite{khalil1996noninear} of a certain region of the state space.
The idea of constraints-driven controls is in stark contrast to finding one optimal trajectory to some specific task.
However,
although there exist converse theorems which
claim that a forward invariant set has a barrier function under certain conditions \cite{ratschan2018converse,wisniewski2016converse,ames2017control},
finding such a set without assuming stability of the system (e.g. \cite{wang2018permissive}) is difficult in general.

Besides, transfer learning is a framework for learning new tasks by exploiting the knowledge already acquired through learning other tasks,
and is related to "lifelong learning" \cite{thrun1995lifelong}.
Our work can be used as a transfer learning technique by regarding a set of good enough policies as useful information shared among other tasks.

In the next section, we present some assumptions together with the notations used in the paper.

\section{Preliminaries}
\label{sec:pre}
Throughout, $\Real$, $\Real_{\geq0}$ and $\integer_{+}$ are
the sets of real numbers, nonnegative real numbers and positive integers,
respectively.
Let $\norm{\cdot}_{\Real^d}:=\sqrt{\inpro{x,x}_{\Real^d}}$ be the norm induced by the inner product $\inpro{x,y}_{\Real^d}:=x^{\T}y$ for $d$-dimensional real vectors $x,y\in\Real^d$, where $(\cdot)^{\T}$ stands for transposition.
Also, let $C^1(\mathcal{A})$ be a class of continuously differentiable function defined over $\mathcal{A}$.
The interior and the boundary of a set $\mathcal{A}$ are denoted by ${\rm int}(\mathcal{A})$ and $\partial\mathcal{A}$, respectively.
In this paper, we consider an agent with system dynamics described by an ordinary differential equation:
\begin{align}
\frac{dx}{dt}=f(x(t)) + g(x(t))u(t), \label{dynamicalsystem}
\end{align}
where $x(t)\in\Real^{n_x}$ and $u(t)\in\U\subset\Real^{n_u}$ are the state and the instantaneous control input of dimensions $n_x,n_u\in\integer_{+}$, $f:\Real^{n_x}\rightarrow\Real^{n_x}$, and $g:\Real^{n_x}\rightarrow\Real^{n_x\x n_u}$.
Let $\mathcal{D}$ be the state space which is an open connected subset of $\Real^{n_x}$, and let $\X\subset\mathcal{D}$ be its compact subset.
The Lie derivatives along $f$ and $g$ are denoted by $L_f$ and $L_g$.
In this work, we make the following assumptions.
\begin{assumption}
	\label{assump1}
	For any locally Lipschitz continuous policy $\phi:\mathcal{D}\rightarrow\U$,
	$f+g\phi$ is locally Lipschitz over $\mathcal{D}$.
\end{assumption}
\begin{assumption}
	\label{assump2}
	The control space $\U(\subset\Real^{n_u})$ is a polyhedron.
\end{assumption}  
With these preliminaries in place, we present the main contribution.
\section{Constraint Learning for Control Tasks}
\label{sec:main}
In this section, we propose {\em limited duration control barrier functions} (LDCBFs), and present their properties and a practical way to find an LDCBF.
\subsection{Limited Duration Control Barrier Functions}
We start this section by the following definition.
\begin{definition}[Limited-duration safety]
	Given an open set of safe states $\mathO\subset\mathcal{D}$, let $\mathcal{C}^T_{\rm LD}$ be a closed nonempty subset of $\mathO$.
	The dynamical system \refeq{dynamicalsystem} is said to be {\em safe up to time $T$},
	if there exists a policy $\phi$ that ensures $x(t)\in\mathO$ for all $0\leq t< T$
	whenever $x(0)\in\mathcal{C}^T_{\rm LD}$.
\end{definition}
Given $B_{\rm LD}:\mathcal{D}\rightarrow\Real_{\geq 0}$ of class $C^1(\mathcal{D})$, let 
\begin{align}
\mathO&:=\left\{x\in\X:B_{\rm LD}(x)< \frac{L}{\beta}\right\}, \;L>0,\;\beta>0,\label{safeset2}\\
&\hspace{-1.5em}\mathcal{C}^T_{\rm LD}=\left\{x\in\X:B_{\rm LD}(x)\leq \frac{Le^{-\beta T}}{\beta}\right\}\subset\mathO,\label{Bset}
\end{align}	
for some $T>0$.
Now, LDCBFs are defined by below.
\begin{definition}[Limited duration control barrier function]
	\label{defLDCBF}
	A function $B_{\rm LD}:\mathcal{D}\rightarrow\Real_{\geq 0}$ of class $C^1(\mathcal{D})$ is called a limited duration control barrier function (LDCBF) for $\mathO$ defined by \refeq{safeset2} and for $T$ if the following conditions are met:
	\begin{enumerate}
	\item $\mathO\subset{\rm int}(\X)$.
	\item $\mathcal{C}^T_{\rm LD}$ defined by \refeq{Bset} is nonempty and there exists a
	monotonically increasing locally Lipschitz continuous function\footnote{Note $\alpha$ is not necessarily an extended class-$\mathcal{K}$ function \cite{khalil1996noninear}.} $\alpha:\Real\rightarrow\Real$ 
	such that $\alpha(0)=0$ and
	\begin{align}
	&\hspace{-2.8em}\inf_{u\in\U}\left\{L_fB_{\rm LD}(x)+L_gB_{\rm LD}(x)u\right\}\nn\\
	&\hspace{-2.8em}\leq\alpha\left(\frac{Le^{-\beta T}}{\beta}-B_{\rm LD}(x)\right)+\beta B_{\rm LD}(x),~\forall x\in\mathO. \nn
	\end{align}	
	\end{enumerate}
\end{definition}
Given an LDCBF, the admissible control space $\mathcal{S}^T_{\rm LD}(x),\;x\in\mathO$, is
defined by
\begin{align}
\mathcal{S}^T_{\rm LD}&(x):=\{u\in\U:L_fB_{\rm LD}(x)+L_gB_{\rm LD}(x)u\nn\\
&\leq\alpha\left(\frac{Le^{-\beta T}}{\beta}-B_{\rm LD}(x)\right)+\beta B_{\rm LD}(x)\}.
\label{admissiblec}
\end{align}
If the initial state is taken in $\mathcal{C}^T_{\rm LD}$ and
an admissible control is employed, safety up to time $T$ is guaranteed.
\begin{theorem}
	\label{Theo1}
	Suppose that a set of safe states $\mathO$ defined by \refeq{safeset2} and an
	LDCBF $B_{\rm LD}$ defined on $\mathcal{D}$ are given.  Suppose also that $x(0)\in\mathcal{C}^T_{\rm LD}$, where $\mathcal{C}^T_{\rm LD}$ is defined by \refeq{Bset}.
	Then, under Assumption \ref{assump1}, any locally Lipschitz continuous
	policy $\phi:\mathcal{D}\rightarrow\U$ that satisfies $\phi(x)\in\mathcal{S}^T_{\rm LD}(x),\;\forall x\in\mathO$, renders the dynamical system \refeq{dynamicalsystem} safe up to time $T$.
\end{theorem}
\begin{proof}
	See Appendix \ref{app:theo1}.
\end{proof}	
In practice, one can constrain the control input within the admissible control space $\mathcal{S}^T_{\rm LD}(x),\;x\in\mathO$, via QPs in the same manner as CBFs and CLFs.
\begin{proposition}
	\label{Pro1}
	Given an LDCBF $B_{\rm LD}$ with a locally Lipschitz derivative and the admissible control space $\mathcal{S}^T_{\rm LD}(x^{*})$ at $x^{*}\in\mathO$ defined by \refeq{admissiblec},
	consider the QP:
	\begin{align}
	\phi(x^{*})=\displaystyle\argmin_{u\in\mathcal{S}^T_{\rm LD}(x)}u^{\T}H(x^{*})u+2b(x^{*})^{\T}u, \nonumber
	\end{align}	
	where $H$ and $b$ are Lipschitz continuous at $x^{*}\in\mathO$, and $H(x^{*})=H^{\T}(x^{*})$ is positive definite.
	If the width\footnote{See Appendix \ref{app:width} for the definition.} of a feasible set is strictly larger than zero, then under Assumption \ref{assump2}, the minimizers $\phi(x^{*})$ are unique and Lipschitz with respect to the state at $x^{*}$.
\end{proposition}	
\begin{proof}
	Slight modifications of {\cite[Theorem~1]{morris2013sufficient}} proves the proposition. 
\end{proof}	
\begin{remark}
	Assumption \ref{assump2} is required for the constraints to be entirely expressed as the intersection of finite affine constraints.
\end{remark}
As such, through LDCBFs, global property (i.e., limited-duration safety) is ensured by constraining instantaneous control inputs.
A benefit of considering LDCBFs is that one can systematically obtain it under mild conditions.
\subsection{Finding a Limited Duration Control Barrier Function}
\label{subsec:finding}
We present a possible way to find an LDCBF $B_{\rm LD}$ for the set of safe states through value function learning.
Here, we should mention that, in practice, one may consider cases where a nominal model or a simulator is available to learn LDCBF during training time, or cases where getting outside of safe regions during training is not "fatal" (e.g., breaking the agent).

Let $\ell:\mathcal{D}\rightarrow\Real_{\geq 0}$, be the immediate cost\footnote{In this paper, we consider the costs that do not depend on control inputs.}, and suppose $\mathO\subset{\rm int}(\X)$ where the set of safe states $\mathO$ is given by
\begin{align}
\mathO:=\left\{x\in\mathcal{D}:\ell(x)< L\right\}, \;L>0.\nn
\end{align}	
Given a policy $\phi:\mathcal{D}\rightarrow\U$, suppose that the system \refeq{dynamicalsystem} is locally Lipschitz and that the initial condition $x(0)=x$ is in $\mathO$.  Then, following the first argument in Appendix \ref{app:theo1}, $x(t)$ can be uniquely defined by extending the solution until reaching $\partial\X$.
Let $T_e(x)$ be the first time at which the trajectory $x(t)$ exits $\mathO$ when $x(0)=x\in\mathO$. 
Now, we define the value function $V^{\phi,\beta}:\mathcal{D}\rightarrow\Real_{\geq 0}$ by
\begin{align}
\hspace{-2em}V^{\phi,\beta}(x):=
\begin{cases}
\int_{0}^{T_e(x)}e^{-\beta t}\ell(x(t))dt+\frac{Le^{-\beta T_e(x)}}{\beta}~~~ (x\in\mathO) \\
\frac{\ell(x)}{\beta}~~~~~~~~~~~~~~~~~~~~~~~~~~~~~~~~~~~~~(x\in\mathcal{D}\setminus\mathO)
\end{cases}\nonumber
\end{align}
where $\beta> 0$ is the discount factor.
When the restriction of $V^{\phi,\beta}$ to $\mathO$, denoted by $V^{\phi,\beta}|_{\mathO}$, is of class $C^1(\mathO)$, we obtain the continuous-time Bellman equation \cite{lewis2009reinforcement}:
\begin{align}
\hspace{-1em}\beta V^{\phi,\beta}(x)=L_fV^{\phi,\beta}(x)+L_gV^{\phi,\beta}(x)&\phi(x)+\ell(x),\nn\\
& \forall x\in\mathO. \label{inf}
\end{align}
Now, for $V^{\phi,0}(x):=\int_{0}^{\infty}\ell(x(t))dt,~x\in\mathO$, to exist and to be a CLF that ensures controlled invariance of its sublevel sets, one must at least assume that the policy $\phi$ stabilizes the agent in a state $x^{*}\in\mathO$ where $\ell(x^{*})=0$ and $\ell(x)>0,~\forall x\in\mathO\setminus\{x^{*}\}$, which is restrictive.  Instead, one can use $V^{\phi,\beta}$ as an LDCBF when $\beta>0$.

Let $\hat{V}^{\phi,\beta}:\mathcal{D}\rightarrow\Real_{\geq0}$ of class $C^1(\mathcal{D})$ denote an approximation of $V^{\phi,\beta}$.
Since $V^{\phi,\beta}(x)\geq\frac{L}{\beta}$ for all $x\in\mathcal{D}\setminus\mathO$ by definition, it follows that
\begin{align}
\left\{x\in\mathcal{D}:V^{\phi,\beta}(x)< \frac{L}{\beta}\right\}\subset\mathO\subset{\rm int}(\X).\nn
\end{align}	
Therefore, we wish to use the approximation $\hat{V}^{\phi,\beta}$ as an LDCBF for the set $\mathO$; however, because it has an approximation error, we take the following steps to guarantee limited-duration safety.
Using \refeq{inf}, define the estimated immediate cost function $\hat{\ell}$ by
\begin{align}
\hspace{-2.0em}\hat{\ell}(x)=\beta \hat{V}^{\phi,\beta}(x)-L_f\hat{V}^{\phi,\beta}(x)-&L_g\hat{V}^{\phi,\beta}(x)\phi(x),\forall x\in\mathO.\nn
\end{align}	
Select $c\geq0$ so that $\hat{\ell}_c(x):=\hat{\ell}(x)+c\geq0$ for all $x\in\mathO$, and define
the function $\hat{V}_c^{\phi,\beta}(x):=\hat{V}^{\phi,\beta}(x)+\frac{c}{\beta}$.
\begin{theorem}
	\label{theoFB}
	Given $T>0$, consider
	the set 
	\begin{align}
	\hat{\mathcal{C}}^{T}_{\rm LD}=\left\{x\in\X:\hat{V}_c^{\phi,\beta}(x)\leq \frac{\hat{L}e^{-\beta T}}{\beta}\right\},\label{hatV}
	\end{align}	
	where $\hat{L}:=\inf_{\displaystyle y\in\X\setminus\mathO}\beta\hat{V}_c^{\phi,\beta}(y)$.
	If $\hat{\mathcal{C}}^{T}_{\rm LD}$ is nonempty, then
	$\hat{V}_c^{\phi,\beta}(x)$ is an LDCBF for $T$ and for the set
	\begin{align}
	\hat{\mathO}:=\left\{x\in\X:\hat{V}_c^{\phi,\beta}(x)< \frac{\hat{L}}{\beta}\right\}\subset\mathO.\nn
	\end{align}	
\end{theorem}
\begin{proof}
	See Appendix \ref{app:theFB}.
\end{proof}
\begin{remark}
	The procedures above basically considers more conservative sets $\hat{\mathcal{C}}^{T}_{\rm LD}$ and $\hat{\mathO}$ so that the approximation error incurred by using $\hat{V}_c^{\phi,\beta}$ is taken into account. 
	One can select sufficiently large $c$ and sufficiently small $\hat{L}$ in practice to make the set of safe states more conservative.
	To enlarge the set $\mathcal{C}^T_{\rm LD}$, the immediate cost $\ell(x)$ is preferred to be close to zero
	for $x\in\mathO$, and $L$ needs to be sufficiently large.  Also, to make $T$ as large as possible, the given policy $\phi$ should keep the system safe up to sufficiently long time (see also Definition \ref{defge} for {\em good enough policy}); given a policy, the larger $T$ one selects the smaller the set $\mathcal{C}^T_{\rm LD}$ becomes.
	In addition, when $\ell(x)$ is almost zero inside $\mathO$ and when $L\gg1$, the choice of $\beta$ does not matter significantly to conservativeness of $\mathcal{C}^T_{\rm LD}$.
\end{remark}

As our approach is set-theoretic rather than specifying a single optimal policy, it is also compatible with the constraints-driven control and transfer learning.
\section{Applications}
\label{sec:app}
In this section, we present two practical applications of LDCBFs, namely, long-duration autonomy and transfer learning.
\subsection{Applications to Long-duration Autonomy}
\label{subsec:app1}
In many applications, guaranteeing particular properties (e.g., forward invariance) over an infinite-time horizon is difficult.
Nevertheless, it is often sufficient to guarantee safety up to certain finite time, and
our proposed LDCBFs act as useful relaxations of CBFs.
To see that one can still achieve long-duration autonomy by using LDCBFs,
we consider the settings of work in \cite{notomista2018persistification}.
\subsubsection{Problem Formulation}
Suppose that the state $x:=[E,p^{\T}]^{\T}\in\Real^3$ has the information of energy level $E\in\Real_{\geq0}$ and the position $p\in\Real^2$ of an agent, and the dynamics is given by \refeq{dynamicalsystem} for
\begin{align}
	f(x)=[\hat{F}(x),F(x)]^{\T},~~~~g(x)=[\hat{G}(x),G(x)]^{\T},\nn
\end{align}
where $\hat{F}:\Real^{3}\rightarrow\Real$, $F:\Real^{3}\rightarrow\Real^{2}$, $\hat{G}:\Real^{3}\rightarrow\Real^{2}$, and $G:\Real^{3}\rightarrow\Real^{2\x2}$.  
Suppose also that the minimum necessary energy level $E_{\rm min}$ and the maximum energy level $E_{\rm max}$ satisfy 
$0< E_{\rm min}<E_{\rm max}$, 
and that $\rho(p)\geq0$ (equality holds only when the agent is at a charging station) is the energy required to bring the agent to a charging station from $p\in\Real^2$, where $\rho\in C^1(\Real^{2})$.
Define $\Delta E := E_{\rm max}-E_{\rm min}$ and
	\begin{align}
	\hspace{-2em}\X:=\{x\in\Real^3:E_{\rm min}\leq E\leq E_{\rm max}\wedge0\leq\rho(p)\leq \Delta E\}.\nn
	\end{align}
Also, let $\U\subset\Real^{2}$ be a control space.
The open connected set $\mathcal{D}\supset\X$ is assumed to be properly chosen.
Then, for $L:=\beta \cdot\Delta E,~\beta>0$, we define $B_{\rm LD}:\mathcal{D}\rightarrow\Real_{\geq 0}$ by
\begin{align}
B_{\rm LD}(x):=\tilde{H}_{\epsilon}(E)+\rho(p),\nn
\end{align}
where $\frac{\Delta E}{4}\gg\epsilon>0$ and $\tilde{H}_{\epsilon}(E)=E_{\rm max}-E$ for $\forall E<E_{\rm max}-3\epsilon$ (see Appendix \ref{app:defhuber} for the detailed definition, which ensures $\mathO\subset{\rm int}(\X)$).
Given $T>0$, we can define $\mathO$ and $\mathcal{C}^T_{\rm LD}$ by \refeq{safeset2} and \refeq{Bset}.
Note $x\in\mathO$ implies $E>E_{\rm min}+\rho(p)$.
\begin{assumption}
	\label{assump3}
	The energy dynamics satisfies
	\begin{align}
	\exists K_d>0,~~~\frac{dE}{dt}=\hat{F}(x)+\hat{G}(x)u&\geq-K_d,\nn\\
	&\forall x\in\mathcal{D},~\forall u\in\U,\nn
	\end{align} and is upper bounded by
	$\frac{dE}{dt}\leq0,~\forall x\in\mathcal{A}_{\rho(p)=0},~\forall u\in\U$, where
	\begin{align}
	\mathcal{A}_{\rho(p)=0}:=\{x\in\mathO:\rho(p)=0\}.\nn
	\end{align}
	In addition, the set
	\begin{align}
	\tilde{\mathcal{S}}^T_{\rm LD}&(x):=\{u\in\U:L_{\tilde{f}}B_{\rm LD}(x)+L_{\tilde{g}}B_{\rm LD}(x)u\nn\\
	&\leq\alpha\left(\frac{Le^{-\beta T}}{\beta}-B_{\rm LD}(x)\right)+\beta B_{\rm LD}(x)\},\label{assump3ineq}
	\end{align}
	is nonempty for all $x\in\mathO\setminus(\mathcal{A}_{\rho(p)=0}\cup\mathcal{A}_{E})$, where
	$\tilde{f}(x)=[-K_d,F(x)]^{\T},~\tilde{g}=[\signal{0},G(x)]^{\T}$, and
	\begin{align}
	\mathcal{A}_{E}:=\{x\in\mathO:E\geq E_{\rm max}-4\epsilon\}.\nn
	\end{align}
\end{assumption}
\begin{remark}
	\label{longremark}
	Suppose $\mathcal{S}^T_{\rm LD}(x)$ is nonempty for all $x\in\mathcal{A}_{\rho(p)=0}\cup\mathcal{A}_{E}$.  Suppose also that Assumption \ref{assump3} holds, then $B_{\rm LD}$ is an LDCBF for $\mathO$ and $T$ s.t. $\mathcal{C}^T_{\rm LD}$ is nonempty, because $\tilde{\mathcal{S}}^T_{\rm LD}(x)\subset\mathcal{S}^T_{\rm LD}(x)$ for all $x\in\mathO\setminus(\mathcal{A}_{\rho(p)=0}\cup\mathcal{A}_{E})$.
\end{remark}
Assumption \ref{assump3} implies that the least possible exit time $\hat{T}_{\rm energy}(E)$ of $E>E_{\rm min}$ being below $E_{\rm min}$ is 
\begin{align}
\hat{T}_{\rm energy}(E)=\frac{(E-E_{\rm min})}{K_d}.\nn
\end{align}
Under these settings, the following proposition holds.
\begin{proposition}
	\label{proconst}
	Suppose Assumption \ref{assump1} and Assumption \ref{assump3} hold, and that $T>\hat{T}_{\rm energy}(E_0)$ for the initial energy level $E_{\rm max}-4\epsilon\geq E_0>E_{\rm min}$.
	Suppose also that $x(0)\in\mathcal{C}^T_{\rm LD}$,
	and that a locally Lipschitz continuous policy $\phi:\mathcal{D}\rightarrow\U$ satisfies 
	$\phi(x)\in\tilde{\mathcal{S}}^T_{\rm LD}(x)$ for all $x\in\mathO\setminus(\mathcal{A}_{\rho(p)=0}\cup\mathcal{A}_{E})$.
	Further, assume the maximum interval of existence of unique solutions $E_t$ and $\rho_t$, namely, the trajectories of $E$ and $\rho(p)$, is given by $[0,T^{*})$ for some $T^{*}>0$.
	Then,
	\begin{align}
	&T_{\rho_t=0}:=\inf\{t\in[0,T^{*}):\rho_t= 0\}\nn\\&\leq T_{\rm energy}:=\inf\{t\in[0,T^{*}):E_t-E_{\rm min}\leq 0\}.\nn
	\end{align}
\end{proposition}
\begin{proof}
	See Appendix \ref{app:proconst}.
\end{proof}
\begin{remark}
	When a function $B_{\rm LD}$ satisfying Assumption \ref{assump3} and the set $\mathO$ are given, we assume that $\tilde{\mathcal{S}}^T_{\rm LD}$ defined for a smaller constant $\beta^{*}<\beta$, instead of $\beta$, is nonempty as well.  Then the agent stays in $\mathO$ longer than $T$ if taking the control input in $\tilde{\mathcal{S}}^T_{\rm LD}$ starting from inside $\mathcal{C}^T_{\rm LD}$, which is defined for $\beta$.  In this case, there is a trade-off between $\beta$ and $T$.
\end{remark}
\begin{remark}
	Instead of assuming \refeq{assump3ineq}, one may learn an LDCBF following the arguments in Section \ref{subsec:finding}.  In such a case, the immediate cost function $\ell(x)$ may be defined so that $0\leq\ell(x)\ll1$ for $E\in(E_{\rm min},E_{\rm max})$ and that $\ell(x)\geq 1$ otherwise.  Then, one may learn the value function of some policy for the system where $\hat{F}(x)=-K_d,~\forall x\in\mathO\setminus\mathcal{A}_{\rho(p)=0}$, $\hat{G}(x)=\signal{0},~\forall x\in\mathO$, and $\exists y\in\mathcal{A}_{\rho(p)=0},~\exists K_i>0,~\hat{F}(y)\geq K_i$.  If $\hat{\mathcal{C}}^T_{\rm LD}$ defined by \refeq{hatV} is nonempty for $T>\hat{T}_{\rm energy}(E_0)$, then similar claims to Proposition \ref{proconst} hold.  Note the policy does not have to stabilize the system around $\mathcal{A}_{\rho(p)=0}$ but can be anything as long as $\hat{\mathcal{C}}^T_{\rm LD}$ becomes nonempty.
\end{remark}
\subsubsection{Simulated Experiment}
Let the parameters be $E_{\rm max}=1.0$, $E_{\rm min}=0.55$, $K_d=0.01$, $\beta=0.005$ and $T=50.0>45.0=\Delta E/K_d$.  We consider six agents (robots) with single integrator dynamics.
An agent of the position $p_i:=[{\rm x}_i,{\rm y}_i]^{\T}$ is assigned a charging station of the position $p_{{\rm charge},i}$, where ${\rm x}$ and ${\rm y}$ are the X position and the Y position, respectively.
When the agent
is close to the station (i.e., $\norm{p_i-p_{{\rm charge},i}}_{\Real^2}\leq0.05$), it remains there until the battery is charged to $E_{\rm ch}=0.92$.
Actual battery dynamics is given by $dE/dt=-0.01E$.
The coverage control task is encoded as Lloyd's algorithm \cite{cortes2017coordinated} aiming at converging to the Centroidal Voronoi Tesselation, but with a soft margin so that the agent prioritizes the safety constraint.
The locational cost used for the coverage control task is given by the following \cite{cortes2004coverage}:
\begin{align}
\sum_{i=1}^6\int_{V_i(p)}\norm{p_i-\hat{p}}^2\varphi(\hat{p})d\hat{p},\nn
\end{align}
where $V_i(p)=\{\hat{p}\in\Real^{2}:\norm{p_i-\hat{p}}\leq\norm{p_j-\hat{p}},\forall j\neq i\}$
is the Voronoi cell for the agent $i$.
In particular, we used $\varphi([\hat{\rm x},\hat{\rm y}]^{\T})=e^{-\left\{(\hat{\rm x}-0.2)^2+(\hat{\rm y}-0.3)^2\right\}/0.06}+ 0.5e^{-\left\{(\hat{\rm x}+0.2)^2+(\hat{\rm y}+0.1)^2\right\}/0.03}$.
In MATLAB simulation (the simulator is provided on the Robotarium \cite{pickem2017robotarium} website: www.robotarium.org), we used the random seed rng(5) for determining the initial states.
Note, for every agent, the energy level and the position are set so that it starts from inside the set
$\mathcal{C}^T_{\rm LD}$.
Note also that battery information is local to each agent who has its own LDCBFs\footnote{We are implicitly assuming that $B_{\rm LD}$ is an LDCBF, i.e., the conditions in Remark \ref{longremark} are satisfied.}; limited-duration safety is thus enforced in a decentralized manner.

\begin{figure}[t]
	\begin{minipage}{\hsize}
		\begin{center}
			\includegraphics[clip,width=0.62\textwidth]{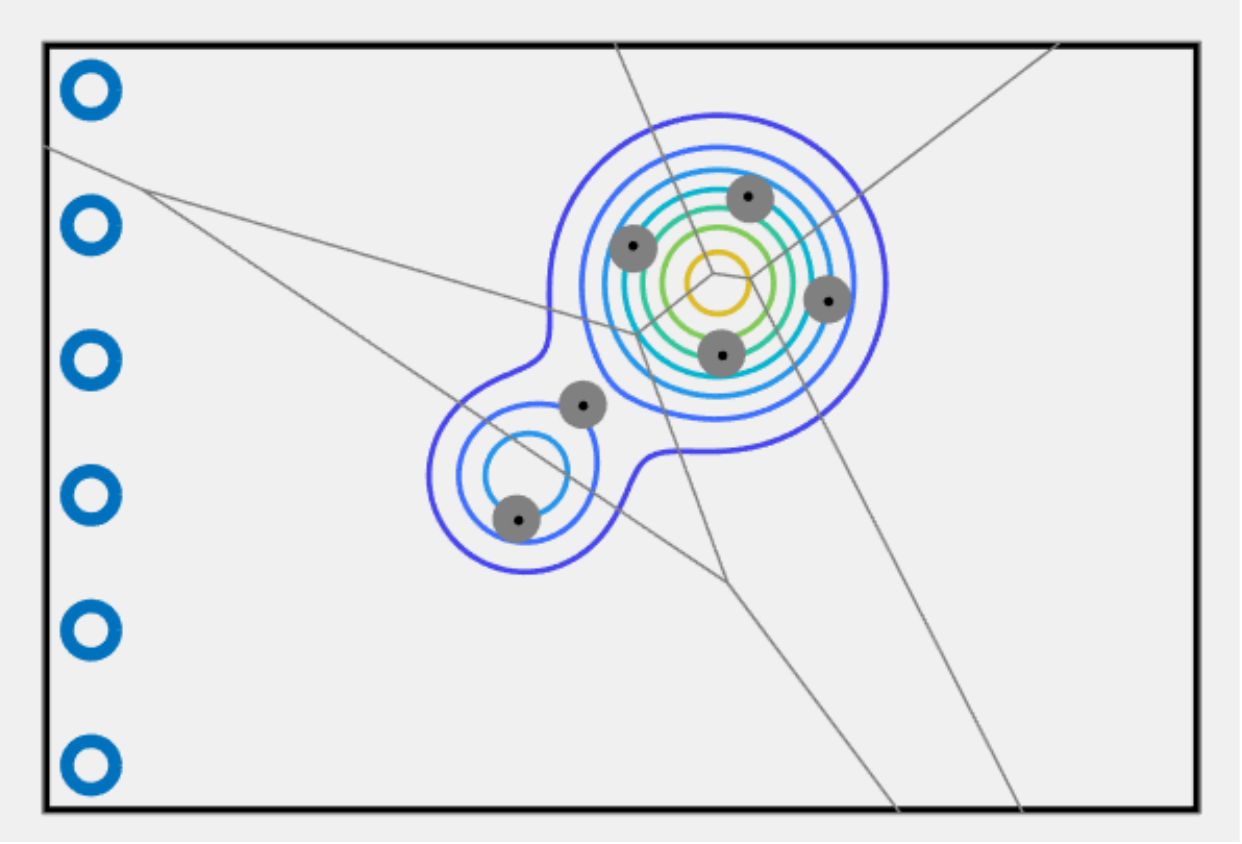}
			\label{fig:robots1}
			\centerline{(a) }
		\end{center}	
	\end{minipage}
	\begin{minipage}{\hsize}
		\begin{center}
			\vspace{1em}
			\includegraphics[clip,width=0.62\textwidth]{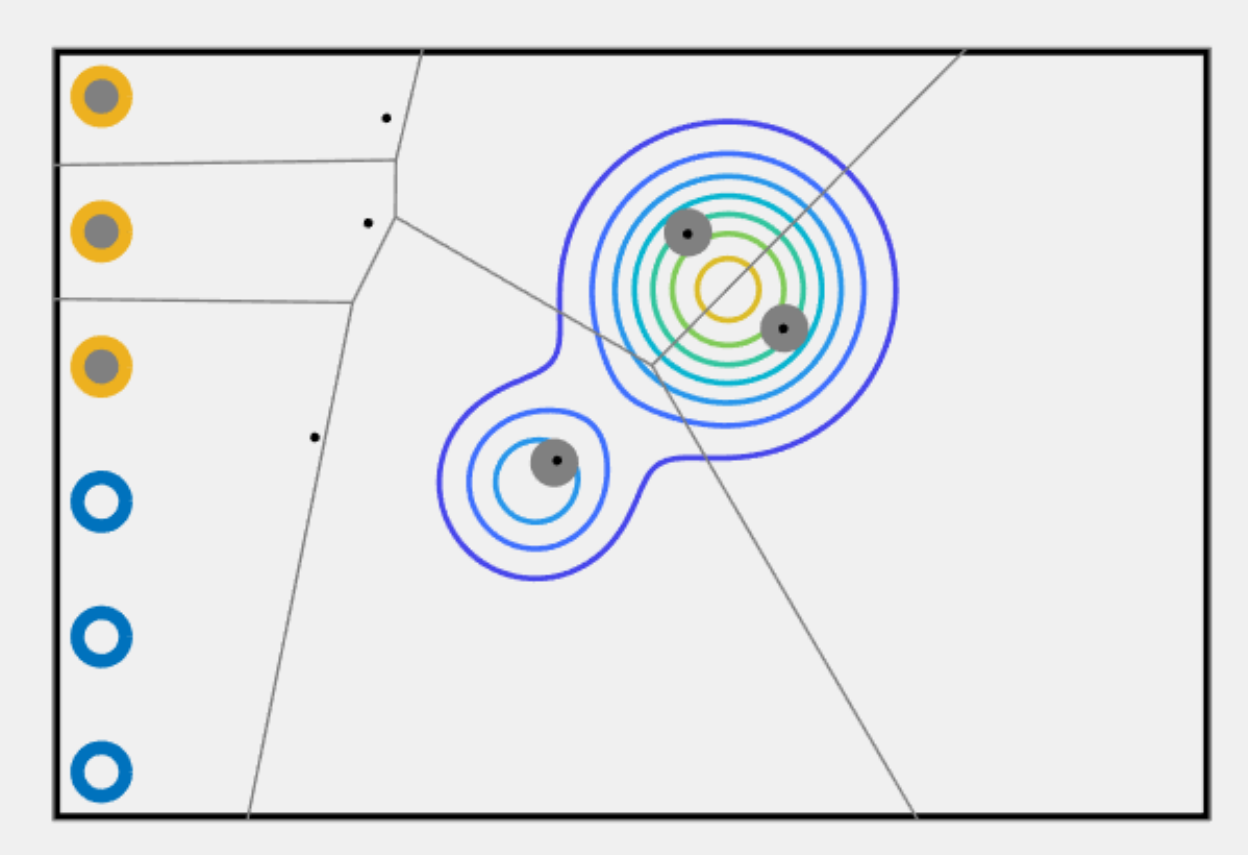}
			\label{fig:robots2}
			\centerline{(b) }
		\end{center}
	\end{minipage}
	\caption{(a) Screenshot of agents executing coverage controls. (b) Screenshot of agents three of which are charging their batteries.}
	\label{fig:expadd}
\end{figure}
\begin{figure}[t]
	\begin{center}
		\includegraphics[clip,width=0.37\textwidth]{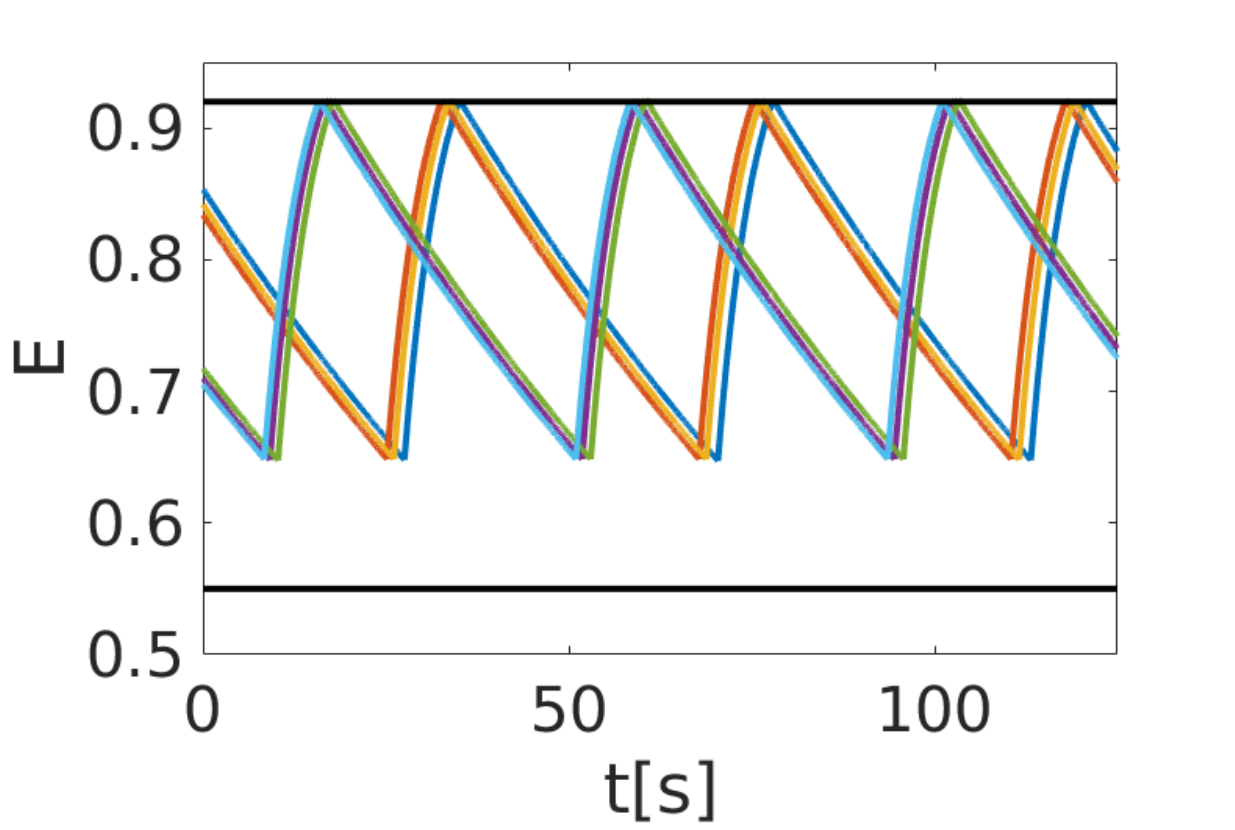}
		\caption{Battery levels of the six agents over time.  Two black lines indicate the energy level when charged ($E_{\rm ch}=0.92$) and the minimum energy level ($E_{\rm min}=0.55$).  All agents successfully executed tasks without depleting their batteries.}
		\label{fig:battery}
	\end{center}	
\end{figure}

\reffig{fig:expadd} shows (a) the images of six agents executing coverage tasks and (b) images of the agents three of which are charging their batteries. 
\reffig{fig:battery} shows the simulated battery voltage data of the six agents, from which we can observe that LDCBFs worked effectively for the swarm of agents to avoid depleting their batteries.

\subsection{Applications to Transfer Learning}
\label{subsec:app2}
Another benefit of using LDCBFs is that, once a set of {\it good enough} policies that guarantee limited-duration safety for sufficiently large $T$ and for sufficiently large $\mathcal{C}^T_{\rm LD}$ is obtained, one can reuse them for different tasks.
\begin{definition}[Good enough policy]
	\label{defge}
	Suppose a policy $\phi$ guarantees safety up to time $T$ if the initial state is in $\mathcal{C}^T_{\rm LD}\subset\mathcal{D}$.
	Suppose also that an initial state of a task $\mathcal{T}$ is always taken in $\mathcal{C}^T_{\rm LD}$ and that the task can be achieved within the time horizon $T$.
	Then, the policy $\phi$ is said to be {\em good enough} with respect to the task $\mathcal{T}$.
\end{definition}
We introduce the definition of transfer learning below.
\begin{definition}[Transfer learning, {\cite[modified version of Definition~1]{pan2010survey}}]
	Given a set of training data $D_S$ for one task (i.e., source task) denoted by
	$\mathcal{T}_S$ (e.g., an MDP) and a set of training data $D_T$ for another task (i.e., target task) denoted by
	$\mathcal{T}_T$, transfer learning aims to improve the learning of the
	target predictive function $f_T$ (i.e., a policy in our example) in $D_T$ using the knowledge in
	$D_S$ and $\mathcal{T}_S$, where $D_S\neq D_T$, or $\mathcal{T}_S\neq\mathcal{T}_T$.
\end{definition}
In our example, we assume we know the state constraints that a target task $\mathcal{T}_T$ is better to satisfy and that some of these constraints are shared with a source task $\mathcal{T}_S$.  If a set of training data $D_S$ is used to obtain a good enough policy for the source task, one can learn an LDCBF with this policy.  When this policy is also good enough for the target task, which would be the case if some constraints are shared with a target task, then the learned LDCBF is expected to be used to speed up the learning of the target task.\footnote{Further study of rigorous sample complexity analysis for this transfer learning framework is beyond the scope of this paper.}
\subsubsection{Illustrative Example}
For example, when learning a good enough policy for the balance task of the cart-pole problem, one can simultaneously learn a set of
limited-duration safe policies that keep the pole from falling down up to certain time $T>0$.
The set of these limited-duration safe policies is obviously useful for other tasks such as moving the cart to one direction without
letting the pole fall down.

We study some practical implementations.
Given an LDCBF $B_{\rm LD}$, define the set $\Phi^T$ of admissible policies as
\begin{align}
\Phi^T:=\{\phi\subset \Phi:\phi(x)\in\mathcal{S}^T_{\rm LD}(x),\;\forall x\in\mathO\},\nn
\end{align}
where $\Phi:=\{\phi:\phi(x)\in\U,\;\forall x\in\mathcal{D}\}$, and $\mathcal{S}^T_{\rm LD}(x)$ is the set of admissible control inputs at $x$.
If an optimal policy $\phi^{\mathcal{T}_T}$ for the target task $\mathcal{T}_T$ is included in $\Phi^T$, one can conduct learning for the target task within the policy space $\Phi^T$.
If not, one can still consider $\Phi^T$ as a soft constraint and explore the policy space $\Phi\setminus\Phi^T$ with
a given probability or one may just select the initial policy from $\Phi^T$.

In practice, a parametrized policy is usually considered; a policy $\phi_{\theta}$ expressed by a parameter $\theta\in\Real^{n_{\theta}}$ for $n_{\theta}\in\integer_{+}$
is updated via policy gradient methods \cite{sutton2000policy}. 
If the policy is in the linear form with a fixed feature vector, the projected policy gradient method \cite{thomas2013projected} can be used.
Given, a set of finite data points $D\subset\X$, the policy $\phi(x)$ is linear with respect to $\theta$ at each $x\in D$ and that LDCBF constraints are affine with respect to $\phi(x)$ at each $x\in D$; therefore, $\tilde{\Phi}_{\theta}^T:=\{\theta\in\Real^{n_{\theta}}:\phi_{\theta}(x)\in\mathcal{S}^T_{\rm LD}(x),\;\forall x\in D\}$ is an intersection of finite affine constraints, which is a polyhedron.
Hence, the projected policy gradient method looks like
$\theta\leftarrow\Gamma[\theta+\lambda \nabla_{\theta} F^{\mathcal{T}_T}(\theta)]
$.  Here, $\Gamma:\Real^{n_{\theta}}\rightarrow\tilde{\Phi}_{\theta}^T$ projects a policy onto $\tilde{\Phi}_{\theta}^T$ and $F^{\mathcal{T}_T}(\theta)$ is the objective function for the target task which is to be maximized.
For the policy not in the linear form, one may update policies based on LDCBFs
by modifying the deep deterministic policy gradient (DDPG) method \cite{lillicrap2015continuous}:
because through LDCBFs, the global property (i.e., limited-duration safety) is ensured by constraining local control inputs, it suffices to add penalty terms to the cost when updating a policy using samples.
For example, one may employ the log-barrier extension proposed in \cite{kervadec2019log}, which is a smooth approximation of the hard indicator function for inequality constraints but is not restricted to feasible points.
\subsubsection{Simulated Experiment}
The simulation environment and the deep learning framework used in this simulated experiment are "Cart-pole" in DeepMind Control Suite and PyTorch \cite{paszke2017automatic}, respectively.
We take the following steps:
\begin{enumerate}
	\item Learn a policy that balances the pole by using DDPG \cite{lillicrap2015continuous} over sufficiently long time horizon.
	\item Learn an LDCBF by using the obtained actor network.
	\item Try a random policy with the learned LDCBF and using a (locally) accurate model to see that LDCBF works reasonably.
	\item With and without the learned LDCBF, learn a policy that moves the cart to left
	without letting the pole fall down, which we refer to as move-the-pole task.
\end{enumerate}
The parameters used for this simulated experiment are summarized in \reftab{sumparam1}.
Here, angle threshold stands for the threshold of $\cos{\psi}$ where $\psi$ is the angle of the pole
from the standing position, and position threshold is the threshold of the cart position $p$.
The angle threshold and the position threshold are used to terminate an episode.
Note that the cart-pole environment of MuJoCo \cite{todorov2012mujoco} xml data in DeepMind Control Suite is modified so that the cart can move between $-3.8$ and $3.8$.
We use prioritized experience replay when learning an LDCBF.
Specifically, we store the positive and the negative data, and sample $4$ data points from the positive one and the remaining $60$ data points from the negative one.
In this simulated experiment, actor, critic and LDCBF networks use ReLU nonlinearities.
The actor network and the LDCBF network consist of two layers of $300\rightarrow200$ units, and the
critic network is of two layers of $400\rightarrow300$ units. The control input vector is concatenated to the state vector from the second critic layer.

\noindent{\bf Step1:}
The average duration (i.e., the first exit time, namely, the time when the pole first falls down) out of $10$ seconds (corresponding to $1000$ time steps), over $10$ trials for the policy learned through the balance task by DDPG was $10$ seconds.

\begin{figure}[t]
	\begin{center}
		\hspace{2em}
		\includegraphics[clip,width=0.43\textwidth]{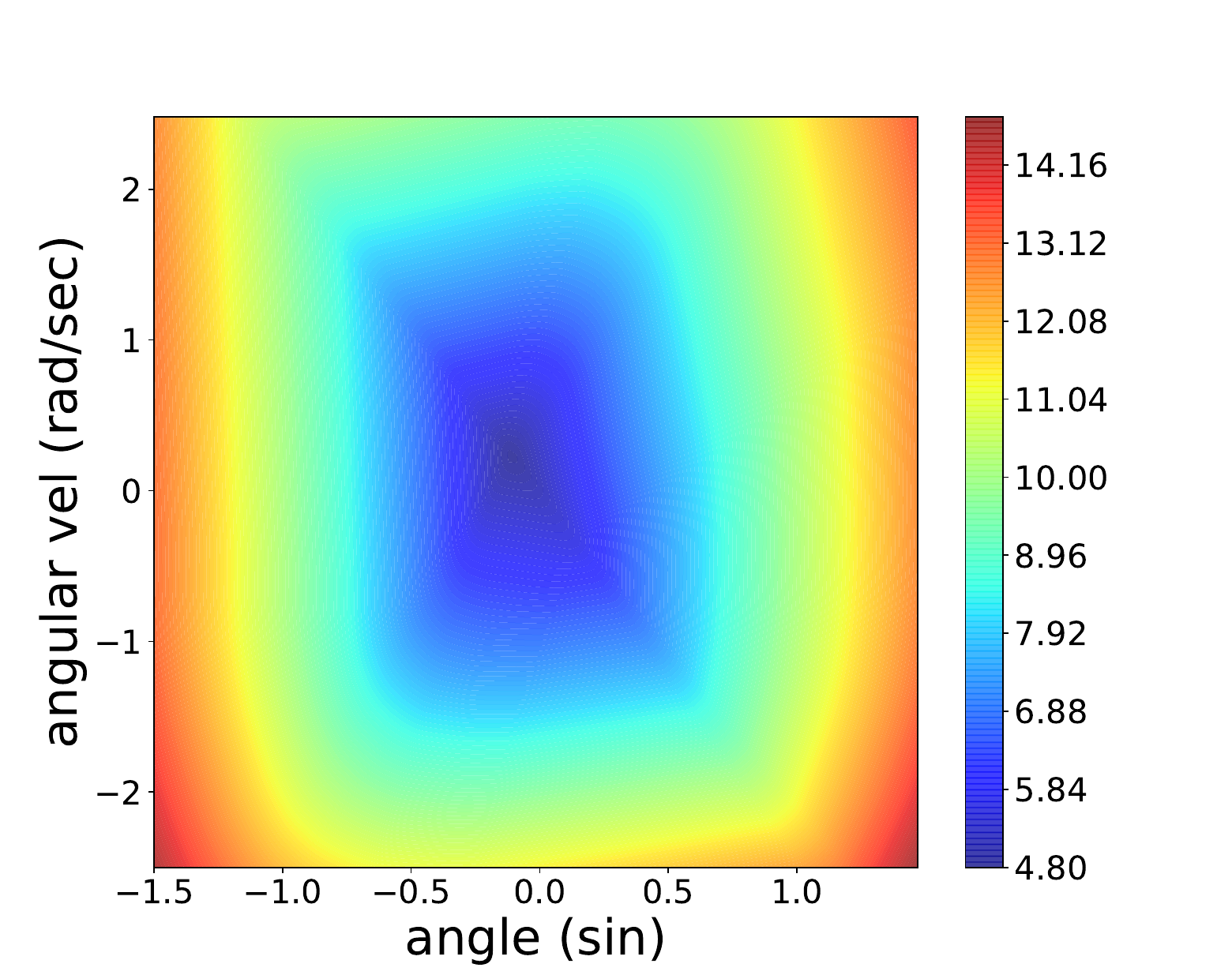}
		\caption{Illustration of the LDCBF for $\sin{\psi}$ and $\dot{\psi}$ at zero cart velocity.
			The center has lower value.  Also, unsafe regions have the values over $\frac{L}{\beta}=\frac{1.0}{-\log{(0.999)}/0.01}\approx10.0$.}
		\label{fig:LDCBFill}
	\end{center}
\end{figure}

\noindent{\bf Step2:}
Then, by using this successfully learned policy, an LDCBF is learned by assigning
the cost $\ell(x)=1.0$ for $\cos{\psi}<0.2$ and $\ell(x)=0.1$ elsewhere.
Also, because the LDCBF is learned in a discrete-time form, we transform it to a continuous-time form
via multiplying it by $\Delta_t=0.01$.
When learning an LDCBF, we initialize each episode as follows:
the angle $\psi$ is uniformly sampled within $-1.5\leq\psi\leq1.5$, the cart velocity $\dot{p}$ is multiplied by $100$ and the angular velocity $\dot{\psi}$ is multiplied by $200$ after being initialized by DeepMind Control Suite.
The LDCBF learned by using this policy is illustrated in \reffig{fig:LDCBFill}, which agrees with our intuitions.  Note that $\frac{L}{\beta}$ in this case is $\frac{1.0}{-\log{(0.999)}/0.01}\approx10.0$.

\noindent{\bf Step3:}
To test this LDCBF, we use a uniformly random policy ($\phi(x)$ takes the value between $-1$ and $1$) constrained by the LDCBF with the function $\alpha(q)=\max{\{0.1q,0\}}$ and with the time constant $T=5.0$.
When imposing constraints, we use the (locally accurate) control-affine model of the cart-pole in the work \cite{barto1983neuronlike},
where we replace the friction parameters by zeros for simplicity.
The average duration out of $10$ seconds over $10$ trials for this random policy was $10$ seconds,
which indicates that the LDCBF worked sufficiently well.
We also tried this LDCBF with the function $\alpha(q)=\max{\{3.0q,0\}}$ and $T=5.0$,
which resulted in the average duration of $5.58$ seconds.
Moreover, we tried the fixed policy $\phi(x)=1.0$, with the function $\alpha(q)=\max{\{0.1q,0\}}$ and $T=5.0$,
and the average duration was $4.73$ seconds, which was sufficiently close to $T=5.0$.

\noindent{\bf Step4:}
For the move-the-pole task,
we define the success by the situation where the cart position $p,\;-3.8\leq p\leq 3.8$, ends up in the region of $p\leq-1.8$ without letting the pole fall down.
The angle $\psi$ is uniformly sampled within $-0.5\leq\psi\leq0.5$ and the rest follow the initialization of DeepMind Control Suite.
The reward is given by
$(1 + \cos{\psi})/2\x$(utils.rewards.tolerance($\dot{p}+1.0$, bounds = ($-2.0,~0.0$), margin = $0.5$)),
where utils.rewards.tolerance is the function defined in \cite{controlsuite}.
In other words, we give high rewards when the cart velocity is negative
and the pole is standing up.
To use the learned LDCBF for DDPG, we store matrices and vectors used in linear constraints along with
other variables such as control inputs and states, which we use for experience replay.
Then, the log-barrier extension cost proposed in \cite{kervadec2019log} is
added when updating policies.
Also, we try DDPG without using the LDCBF for the move-the-pole task.
Both approaches initialize the policy by the one obtained after the balance task.
The average success rates of the policies obtained after the numbers of episodes up to $15$ over $10$ trials are given in \reftab{sumres} for DDPG with the learned LDCBF and DDPG without LDCBF.
This result implies that our proposed approach successfully transferred information from the source task to the target task.

\begin{table*}[t]
	\caption{Summary of the parameter settings for the cart-pole problem.  These parameters are chosen so that a policy for the balance task can be obtained, an LDCBF can be learned, and the move-the-pole task can be accomplished within 15 episodes when using an LDCBF.  We could not find parameters that make the move-the-pole task work without LDCBFs within 15 episodes.}
	\label{sumparam1}
	\centering
	\begin{tabular}{lllll}
		\toprule
		Parameters & Balance task & For Learning & Move-the-pole task  &  Move-the-pole task \\
		&&an LDCBF&with LDCBF&without LDCBF\\
		\midrule
		Discount $\beta$ & $-\log{(0.99)}/0.01$ & $-\log{(0.999)}/0.01$ & $-\log{(0.999)}/0.01$ & $-\log{(0.999)}/0.01$\\
		Angle threshold $\cos{\psi_{\rm thre}}$ & 0.75 & 0.2 & 0.75 & 0.75\\
		Position threshold $p_{\rm thre}$ & $\pm$1.8  & $\pm$3.8 & $\pm$3.8 & $\pm$3.8 \\
		Soft-update $\mu$ & $10^{-3}$ & $10^{-2}$ & $10^{-3}$ & $10^{-3}$ \\
		Step size for target NNs & $10^{-4}$ & $10^{-2}$ & $10^{-4}$ & $10^{-4}$ \\
		Time steps per episode & 300 & 50 & 300 & 300 \\
		Number of episodes & 80 & 200 & Up to 15 & Up to 15 \\
		Minibatch size & 64 & 64 & 64 & 64 \\
		Random seed & 10 & 10 & 10 & 10 \\
		States $x$ & $\sin{\psi},~0.1\dot{p},~0.1\dot{\psi}$ &$\sin{\psi},~\dot{p},~\dot{\psi}$&
		$\sin{\psi},~\dot{p},~\dot{\psi}$ &$\sin{\psi},~\dot{p},~\dot{\psi}$\\
		\bottomrule
	\end{tabular}
\end{table*}

\begin{table*}[t]
	\caption{Summary of the results for the move-the-pole task.  Over 15 episodes, the success rates over 10 trials are shown.  DDPG with LDCBF uses learned LDCBF to constrain control inputs when updating policies by DDPG, and DDPG without LDCBF does not use LDCBFs.  By using LDCBFs the move-the-pole task is shown to be easily learned.}
	\label{sumres}
	\centering
	\begin{tabular}{llllllllllllllll}
		\toprule
		Algorithm\textbackslash Episode & 1  & 2&3&4&5&6&7&8&9&10&11&12&13&14&15 \\
		\midrule
		DDPG with LDCBF & 0.0 & 0.0 & 0.0 &0.4 &0.7 &0.8&1.0&1.0&1.0&0.7&0.7&1.0&1.0&1.0&1.0\\
		DDPG without LDCBF & 0.0 &0.0&0.0&0.0&0.0&0.0&0.0&0.0&0.0&0.0&0.0&0.0&0.0&0.0&0.0\\
		\bottomrule
	\end{tabular}
\end{table*}
\section{Conclusion}
In this paper, we presented a notion of limited-duration safety as a relaxation
of forward invariance of a set of safe states.
Then, we proposed limited-duration control barrier functions to guarantee limited-duration safety
by using agent dynamics.
We showed that LDCBFs can be obtained through value function learning,
and analyzed some of their properties.
LDCBFs were validated through persistent coverage control tasks and were successfully applied to a transfer learning
problem by sharing a common state constraint.
\begin{ack}                          
M. Ohnishi thanks Kai Koike at Kyoto University for valuable discussions on differential equations.
The authors thank the anonymous reviewers for their careful and constructive comments that helped us improve this work.
This work of M. Ohnishi was supported in part by Funai Overseas Scholarship and Wissner-Slivka Endowed Graduate Fellowship.  This work of G. Notomista and M. Egerstedt was supported by the US Army Research Lab through Grant No. DCIST CRA W911NF-17-2-0181.
This work of M. Sugiyama was supported by KAKENHI 17H00757. 
\end{ack}

\appendix
\numberwithin{equation}{section}
\numberwithin{proposition}{section}
\numberwithin{theorem}{section}
\numberwithin{remark}{section}
\numberwithin{assumption}{section}
\renewcommand{\theequation}{\thesection.\arabic{equation}}
\renewcommand{\theproposition}{\thesection.\arabic{proposition}}
\section{Proof of Theorem \ref{Theo1}}
\label{app:theo1}
Under Assumption \ref{assump1},
the trajectories $x(t)$ with an initial condition $x(0)\in\mathcal{C}^T_{\rm LD}\subset\mathcal{D}$ exist and are unique over $0\leq t\leq\delta$ for some $\delta>0$.  Let $[0,T^{*}),~T^{*}>0$, be its maximum interval of existence ($T^{*}$ can be $\infty$), and
let $T_e$ be the first time at which the trajectory $x(t)$ exits $\mathO$, i.e.,
\begin{align}
T_e:=\inf\{t\in[0,T^{*}):x(t)\notin\mathO\}.\label{defTe}
\end{align} 
Because $B_{\rm LD}\in C^1(\mathcal{D})$ and $\mathO\subset{\rm int}(\X)$ imply $\mathO$ is open, it follows that $T_e>0$.
If $T^{*}$ is finite, it must be the case that $x(t^{*})\notin\X$ for some $t^{*}\in[0,T^{*})$ which implies $0\leq t\leq T_{\X}<T^{*}$, where
\begin{align}
T_{\X}:=\inf\{t\in[0,T^{*}):x(t)\in\partial \X\}.\nn
\end{align} 
In this case, because $\mathO\subset{\rm int}(\X)$, it follows that $0<T_e\leq T_{\X}<T^{*}$.
If, on the other hand, $T^{*}=\infty$, then $T_e$ can still be defined by \refeq{defTe}, and either $T_e=\infty$ or $T_e<T^{*}$ hold.
When $T_e=\infty$, it is straightforward to prove the claim; therefore, we focus on the case where $T_e$ is finite and $T_e<T^{*}$.
Let $T_p$ denote
the last time at which the trajectory $x(t)$ passes through the boundary of $\mathcal{C}^T_{\rm LD}$ from inside before first exiting $\mathO$, i.e.,
\begin{align}
T_p:=\sup{\left\{t\in[0,T_e):x(t)\in\partial \mathcal{C}^T_{\rm LD}\right\}}.\nn
\end{align}
Because $\mathcal{C}^T_{\rm LD}$ is closed subset of the open set $\mathO$ and because $x(0)\in\mathcal{C}^T_{\rm LD}$, by continuity of the solution, it follows that $0\leq T_p<T_e$.
Now, the solution to 
\begin{align}
\dot{s}(t)=\beta s(t), \nn
\end{align}	
where the initial condition is given by $s(T_p)=B_{\rm LD}(x(T_p))=\frac{Le^{-\beta T}}{\beta}$, is
\begin{align}
s(t)=B_{\rm LD}(x(T_p))e^{\beta (t-T_p)}, \;\forall t\geq T_p.\nn
\end{align}
It thus follows that
\begin{align}
s(T_p+T)=\frac{L}{\beta}e^{-\beta T}e^{\beta T}=\frac{L}{\beta},\nn
\end{align}
and $T_p+T$ is the first time at which the trajectory $s(t),\;t\geq T_p$, reaches $\frac{L}{\beta}$.

Because $\alpha(\frac{Le^{-\beta T}}{\beta}-B_{\rm LD}(t))\leq0,~\forall t\in[T_p,T_e)$, 
and $\phi(x)\in\mathcal{S}^T_{\rm LD}(x),\;\forall x\in\mathO$, we obtain, by the Comparison Lemma \cite{khalil1996noninear}, {\cite[Theorem~1.10.2]{lakshmikantham1969differential}} and by continuity of the solutions over $t\in[0,T_e]$, that $B_{\rm LD}(x(t))\leq s(t),~\forall t\in[T_p,T_e]$.
If we assume $T_e<T_p+T$, it follows that $B_{\rm LD}(x(T_e))\leq s(T_e)<s(T_p+T)=\frac{L}{\beta}$ which is a contradiction because $\mathO\in{\rm int}(\X)$ and $B_{\rm LD}\in C^1(\mathcal{D})$ imply $B_{\rm LD}(x(T_e))=\frac{L}{\beta}$.  Hence,
$T_e\geq T_p+T$, which proves the Theorem.

\section{On Proposition \ref{Pro1}}
\label{app:width} 
The width of a feasible set is defined by the unique solution to the following linear program:
\begin{align}
\omega^{*}(x) = &\displaystyle\max_{[u^{\T},\omega]^{\T}\in\Real^{n_u+1}} \omega \label{width}\\
{\rm s.t.}&\;\;L_fB_{\rm LD}(x)+L_gB_{\rm LD}(x)u+\omega\nn\\
&~~~~~~~\leq\alpha\left(\frac{Le^{-\beta T}}{\beta}-B_{\rm LD}(x)\right)+\beta B_{\rm LD}(x) \nn\\
&\;\; u+[\omega,\omega\ldots,\omega]^{\T}\in\U\nn
\end{align}	
\section{Proof of Theorem \ref{theoFB}}
\label{app:theFB}
Because, by definition,
\begin{align}
\hat{V}_c^{\phi,\beta}(x)\geq\frac{\hat{L}}{\beta},\;\forall x\in\X\setminus\mathO,\nn
\end{align}	
it follows that 
\begin{align}
\hat{\mathO}=\left\{x\in\X:\hat{V}_c^{\phi,\beta}(x)< \frac{\hat{L}}{\beta}\right\}\subset\mathO.\nn
\end{align}	
Because $\hat{V}_c^{\phi,\beta}\in C^1(\mathcal{D})$ satisfies 
\begin{align}
L_f\hat{V}_c^{\phi,\beta}(x)&+L_g\hat{V}_c^{\phi,\beta}(x)\phi(x)\nn\\
&=\beta \hat{V}_c^{\phi,\beta}(x)-\hat{\ell}_c(x),\;\forall x\in\mathO,\nn
\end{align}
and $\hat{\ell}_c(x)\geq0,\;\forall x\in\mathO$, it follows that
\begin{align}
&L_f\hat{V}_c^{\phi,\beta}(x)+L_g\hat{V}_c^{\phi,\beta}(x)\phi(x)\nn\\
&~~~~~~\leq\alpha\left(\frac{\hat{L}e^{-\beta T}}{\beta}-\hat{V}_c^{\phi,\beta}(x)\right)+\beta\hat{V}_c^{\phi,\beta}(x),\nn
\end{align}
for all $x\in\mathO$ and
for a monotonically increasing locally Lipschitz continuous function $\alpha$ such that $\alpha(q)=0,\;\forall q\leq0$.
Therefore, $\phi(x)\in\U,~\forall x\in\mathO\subset{\rm int}(\X)$ and $\hat{\mathcal{C}}^{T}_{\rm LD}\neq\emptyset$, where $\emptyset$ is the empty set, imply that $\hat{V}_c^{\phi,\beta}$ is an LDCBF for the set $\hat{\mathO}$ and for $T$.
\begin{remark}
	A sufficiently large constant $c$ could be chosen in practice. 
	If $\hat{\ell}_c(x)>0$ for all $x\in\mathO$ and the value function is learned by using a policy $\phi$ such that
	$\phi(x)+[c_{\phi},c_{\phi}\ldots,c_{\phi}]^{\T}\in\U$ for some $c_{\phi}>0$, then the unique solution to the linear program \refeq{width}
	satisfies $\omega^{*}(x)>0,~\forall x\in\mathO$.
\end{remark}
\section{Definition of the function $\tilde{H}_{\epsilon}$}
\label{app:defhuber}
For brevity, let $\bar{E}:=E_{\rm max}-E$ and $\bar{E}_{2\epsilon}:=E-E_{\rm max}+2\epsilon$.
Here, we give the definition of $\tilde{H}_{\epsilon}$ which is not practically relevant but is only required to
make $\mathO\subset{\rm int}(\X)$:
\begin{align}
&\tilde{H}_{\epsilon}(E)\nn\\
&:=\begin{cases}
\bar{E}  &(\bar{E}_{2\epsilon}<-\epsilon)\\
\frac{\Delta E}{8\epsilon^2}(\bar{E}_{2\epsilon}^2+2\epsilon \bar{E}_{2\epsilon}+\epsilon^2)+\bar{E}
&(|\bar{E}_{2\epsilon}|\leq \epsilon)\\
\frac{\Delta E}{2\epsilon}\bar{E}_{2\epsilon}+\bar{E} & (\epsilon<\bar{E}_{2\epsilon})
\end{cases}\nn
\end{align}
Note $\tilde{H}_{\epsilon}(E)=\bar{E}+\frac{\Delta E}{4\epsilon^2}H_{\epsilon}(E_{2\epsilon})$,
where $H_{\epsilon}:\Real\rightarrow\Real$ is the Huber function.
It is straightforward to see that $E=E_{\rm max}\Longrightarrow x\notin\mathO$, which is necessary to make $\mathO\subset{\rm int}(\X)$.
\section{Proof of proposition \ref{proconst}}
\label{app:proconst}
Under Assumption \ref{assump1}, the trajectories $x(t)$ with an initial condition $x(0)\in\mathcal{C}^T_{\rm LD}$, and hence $E_t$ and $\rho_t$, exist and are unique over $0\leq t\leq\delta$ for some $\delta>0$.  Let $[0,T^{*})$ be its maximum interval of existence ($T^{*}$ can be $\infty$), which indeed exists.
Following the same argument as in Appendix \ref{app:theo1}, we only focus on the case where $T_e$ defined by \refeq{defTe} is finite and $0<T_e<T^{*}$ (note $T_e=\infty\Longrightarrow T_{\rm energy}=\infty$ and, in this case, the claim is trivially validated.)
Also, if $E_t-E_{\rm min}>0$ for all $t\in[0,T^{*})$, then $T_{\rm energy}=\inf\emptyset=\infty$, and the claim is trivially validated again.
Therefore, we assume that $T_{\rm energy}<T^{*}$.
Further, define 
\begin{align}
\hat{T}&:=\inf{\{t\in[0,T^{*}):\rho_t=0\land x(t)\notin\mathcal{C}^T_{\rm LD}\}},\nn\\
\hat{T}_e&:=\min{\left\{T_e,\hat{T}\right\}}\leq T_{\rm energy},\nn\\
T_p&:=\sup{\left\{t\in[0,T_e):x(t)\in\partial \mathcal{C}^T_{\rm LD} \right\}}.\nn
\end{align}
Following the same argument as in Appendix \ref{app:theo1}, we have $0\leq T_p<T_e$.
Because it must be the case that $E_t>E_{\rm min},~\forall t\in[0,T_p]$,
we should only consider the case where $\rho_t>0,~\forall t\in[0,T_p]$.
If we assume $\hat{T}_e=\hat{T}$, we obtain $\hat{T}\leq T_e\leq T_{\rm energy}$ which proves the claim.
Therefore, we assume $\hat{T}_e=T_e$.

Let $\hat{E}_t$ be the trajectory following the virtual battery dynamics $d\hat{E}/dt=-K_d$ with the initial condition $\hat{E}_{T_p}=E_{T_p}$, and let $s(t)$ be the unique solution to
\begin{align}
\dot{s}(t)=\beta s(t),~~~t\geq T_p,\nn
\end{align}
where $s(T_p)=B_{\rm LD}(x(T_p))=\Delta Ee^{-\beta T}$.
Also, let $\varrho(t)=s(t)+\hat{E}_t-E_{\rm max},~t\geq T_p$.
Then, the time at which $s(t)$ reaches $\Delta E$ is $T_p+T$ because
$s(T+T_p)=B_{\rm LD}(x(T_p))e^{\beta (T+T_p-T_p)}=\Delta Ee^{-\beta T}e^{\beta (T+T_p-T_p)}=\Delta E$.
Since we assumed $x_t\notin\mathcal{A}_{\rho(p)=0},~\forall t\in[0,T_p]$, under Assumption \ref{assump3}, we have $\hat{T}_{\rm energy}(E_{T_p})\leq\hat{T}_{\rm energy}(E_0)<T$.
Further, we have
\begin{align}
\hspace{-2em}\varrho(t)=B_{\rm LD}(x(T_p))e^{\beta(t-T_p)}+\hat{E}_{T_p}-K_d(t-T_p)-E_{\rm max}.\nn
\end{align}
Hence, we obtain 
\begin{align}
\hat{T}_0:=\inf{\{t\geq T_p:\varrho(t)=0\}}\leq T_p+\hat{T}_{\rm energy}(E_{T_p}).\nn
\end{align}
On the other hand, under Assumption \ref{assump1}, the actual battery dynamics can be written as $dE/dt=-K_d+\Delta(x)$, where $\Delta(x)\geq0$.  Also, because $E_0\leq E_{\rm max}-4\epsilon$, it follows that $E_t\leq E_{\rm max}-4\epsilon$ for all $t\in[0,T^{*})$ under Assumption \ref{assump3}, implying $B_{\rm LD}(x(t))=E_{\rm max}-E_t+\rho_t,~\forall t\in[0,T^{*})$.  Therefore,
$\phi(x)\in\tilde{\mathcal{S}}^T_{\rm LD}(x),~\forall x\in\mathO\setminus(\mathcal{A}_{\rho(p)=0}\cup\mathcal{A}_{E})$, indicates
\begin{align}
\hspace{-1.5em}\frac{dB_{\rm LD}(x(t))}{dt}\leq\beta B_{\rm LD}(x(t))-\Delta(x(t)),~~\forall t\in[T_p,T_e). \nn
\end{align}
Then, because
\begin{align}
&\hspace{-1.5em}\frac{d\left(B_{\rm LD}(x(t))-s(t)\right)}{dt}\leq\beta \left(B_{\rm LD}(x(t))-s(t)\right)-\Delta(x(t))\nn\\
&\hspace{4em}\leq\beta\left(B_{\rm LD}(x(t))-s(t)\right),~\forall t\in[T_p,T_e),\nn
\end{align}
and $\beta\left(B_{\rm LD}(x(T_p))-s(T_p)\right)=0$, we obtain $B_{\rm LD}(x(t))-s(t)\leq-\int_0^t\Delta(x(t))dt,~\forall t\in[T_p,T_e)$.
Here, following the same arguments as Appendix \ref{app:theo1}, we have that $T_e\geq T_p+T> T_p+\hat{T}_{\rm energy}(E_{T_p})$.
Further, it follows that
\begin{align}
&\hspace{-1.7em}\rho_t-\varrho(t)=B_{\rm LD}(x(T_p))-s(t)+E_t-\hat{E}_t\nn\\
&\hspace{-1.7em}\leq-\int_0^t\Delta(x(t))dt+\int_0^t\Delta(x(t))dt=0,~\forall t\in[T_p,T_e),\nn
\end{align}
which, by continuity of the solutions, leads to the inequality $\rho_t\leq\varrho(t),~\forall t\in [T_p,T_e]$.
Hence, we conclude that
\begin{align}
\hat{T}\leq \hat{T}_0\leq T_p+\hat{T}_{\rm energy}(E_{T_p})< T_e.\nn
\end{align}
This is a contradiction to the assumption $\hat{T}_e=T_e$, from which the proposition is proved.

\bibliographystyle{plain}        


\end{document}